# Generating a Map of Well-being Regions using Multiscale Moving Direction Entropy on Mobile Sensors

Yukio Ohsawa , Sae Kondo, Yi Sun and Kaira Sekiguchi

*The University of Tokyo, 7-3-1 Hongo, Bunkyo-ku, Tokyo, Japan*

**Abstract**
The well-being of individuals in a crowd is interpreted as a product of the crossover of individuals from heterogeneous communities, which may occur via interactions with other crowds. The index moving-direction entropy corresponding to the diversity of the moving directions of individuals is introduced to represent such an inter-community crossover. Multi-scale moving direction entropies, composed of various geographical mesh sizes to compute the index values, are used to capture the information flow owing to human movements from/to various crowds. The generated map of high values of multiscale moving direction entropy is shown to coincide significantly with the preference of people to live in each region.

**Keywords**
Mobile-sensor data, multi-scale diversity, well-being, community, crowd

## 1. Introduction

The five dimensions of social well-being have been considered positive emotions, engagement, good relationships, meaning and purpose, and sense of achievement [1]. On the other hand, career, social, financial, physical, and community well-being have been defined as five factors of well-being [2]. Social well-being is linked to the well-being of individuals in such a way that an individual seeks not only freedom, but also norms linked to knowledge, ethics, and constraints accepted by society for living safely.

Supposing that society includes various crowds that involve and affect individuals, this study focuses on the dynamics of crowds with respect to norms created within a crowd via its interaction with other crowds or in communities from which individuals in a crowd may come. Here, we distinguish a crowd from a community in which a community is a group of people sharing a common interest or characteristic but those in a crowd may not. On the other hand, a crowd is located in a region with a special location, but those in a community may meet online. We consider the following two factors when analyzing crowd dynamics:
(1) Intra-crowd contacts of individuals, within the same crowd
(2) Iter-crowd contacts of individuals, bringing norms from different crowds

The contacts of type (1) correspond to such phenomena as explained in crowd-psychology [3,4,5]: In [3,4], the negative effect on the quality of shared information such as biased convergence and deindividuation, originating from the unity caused by the limited group, has been pointed out. On the other hand, in [5,6], positive effects were studied: new norms emerge from communication in a crowd because individuals in a crowd may be from other crowds, which means that a crowd itself has an inter-crowd nature, as in (2). The crossover of heterogeneous ideas carried in from other crowds or originating from communities fosters ideas to emerge [7-10] even within a crowd.

Thus, we expect that biased convergence and deindividuation, bound in (1), are released by the emergent norm due to (2). In this study, we hypothesize that activities in which heterogeneous communities interact in good social relationships, in a crowd where inter-crowd contact occurs, are a core factor of well-being. Such an activity has community well-being by nature and also creates financial well-being for those who gather in a crowd to sell or buy valued items and services created in various communities and imported from other crowds. Regarding human resources as a type of valued item to be evaluated on norms created via the interaction of heterogeneous communities, such an activity also enhances career well-being.

Here, using human mobility data available from smartphone sensors, we computed the values of an index representing the diversity of directions in which individuals move from/to local regions. As illustrated in Figure 1, the movements from/to regions are regarded as the physical dimensions of inter-crowd contacts. A map of regions with high values of this index is the output of this study, which implies the enhancement of inter-crowd activities. Thus, we addressed the problem of generating a map of the regions of well-being via the generated map.

In Section 2, we outline our analysis model of crowd dynamics in a region surrounded by other regions of various levels of inter-crowd activities, bringing the aspect of the multiscale structure of social activities. In Section 3, the index of multiscale Moving Direction Entropy (MDE) corresponding to this hierarchical structure is introduced, which can be computed using mobile sensors in smartphones. The correlation of the MDE with the number of people preferring to live in the region is shown in Section 4 for four scales (mesh sizes of regions) for computing the MDE. In Section 5, we discuss the possibility of mixing multiple layers corresponding to multiscale MDE maps as a method for generating a city map of high-well-being crowds. The maps obtained by the proposed method were evaluated based on their correlation with people's preferences to live in local regions.

## 2. The outline: capturing crowd dynamics on mobility divergence

In the following analysis, we introduce a crowd dynamics model in a region surrounded by other regions that also embrace crowds. We modeled such a relationship among regions as a macro-micro interaction of crowds via a hierarchical social structure, which can be captured by multi-scale analysis of human mobility. Here, in order to capture the social structure, we invent a multi-scale index quantifying the diversity of people moving within each region and across regions that are expected to urge inter-crowd contacts because we associate a crowd with spatial constraints.

The index introduced in Section 3 is computed for small meshes, as shown in the central thick frame in Figure 1, and for large meshes, as shown in the larger thick frame. Given our aim to capture the interaction between crowds, as depicted by the dotted arrows in Figure 1, which show the movements of individuals, the index for the larger region should reflect the effect of adjacency among smaller regions.

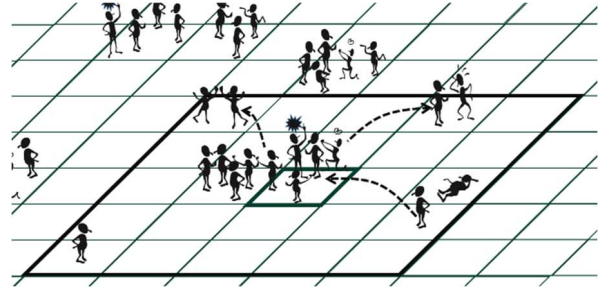

Figure 1 A hierarchical study approach to crowd dynamics. The multi-scale index for a larger mesh shown by the larger thick frame and for the belonging small meshes are computed, and combined to study the macro-micro interaction.

## 3. Multi-scale Moving Direction Entropy
### 3.1. Moving Direction Entropy

As an index for estimating the frequency of inter-crowd contacts, we defined the MDE in region $r$ at time $t$ as in Eq.(1), where $\theta$ denotes the anti-clockwise moving direction, 0 [rad] for north, and $\pi$ for south. $\theta$ is discretized by segmenting $2\pi$ into 100 segments of $\pi/50$ each [10]. $p_\theta(r, t)$ is the probability that an individual moves in the [$\theta$, $\theta + \pi/50$] direction in region $r$ at time $t$.

$$H_{\text{MDE}}(r, t) = \sum_{i=1}^{100} p_{\theta=i\pi/50}(r, t) \log p_\theta(r, t) \quad (1)$$

Eq.(1) represents a type of mobility diversity that concerns the dynamic interaction of various crowds via individual movements from/to various locations corresponding to various crowds. The value of $H_{\text{MDE}}(r, t)$ for each region $r$ and time $t$ is computed from the point data on human mobility and Agoop Inc. (e.g., 615,000 unique smartphone users were monitored at $31 \times 10^6$ logs per month in August 2020). The location and velocity were sensed using a GPS sensor embedded in the smartphone.

Thus far, MDE has been introduced as a quantity that may correlate with inter-community activities when a community and crowd are not distinguished [11]. In addition, MDE is strongly positively correlated with the number of infection cases in the target region [12] because infectious diseases spread widely if people attend multiple communities with comparable priorities [13].

## 3.2. Muti-scale MDE and the results

A region of a mesh sized as the square of length $\Delta$ (100m, 1 km, 2 km, or 4 km) was selected as *r*. A small mesh, as a part of a larger mesh corresponding to the wider thick frame in Figure 1, corresponds to a narrower thick frame. Note that the MDE in the larger mesh is equal to the average of the MDEs in its included small meshes only if there is no same segment of discretized $\theta$ (see [14] for proof where a segment in this study corresponds to a cluster of data samples). Because this condition hardly holds, the value of MDE for a large mesh reflects the similarity of human-movement directions between included small meshes, which implies the movement of people across adjacent local regions. This means that MDE is a suitable index for considering the hierarchical dynamics of crowds. However, for the same reason, the MDE for different mesh scales ($\Delta$'s) cannot be compared. Thus, MDE values are compared within each scale, that is, between meshes of the same size, in the analyses below.

Figure 2 shows the computing results for the area of (E139.3-140.0, N35.5-35.85) in Tokyo. The peaks of the red heatmap showing high-MDE regions fit the blue dots showing the highly preferred residential regions, many of which are clustered in the central region, that is, the most activated part of Tokyo. Here, the preferred stations, such as Hachioji, Chofu, Kasai, and Chitose Karasuyama, are not covered by high-MDE meshes for $\Delta$= 100m, whereas they are covered for larger $\Delta$.

On the other hand, Figure 3, by the *y* axis, shows the number of near-station regions, that are among the 43 most highly preferred to live in, within *x* km from the highest-MDE grids (centers of meshes) of each mesh size $\Delta$. In other words, Figure 3 shows the *recall* of preferred residential regions given by the stations to live nearby using MDE computation. Here, we find that the highest granularity, $\Delta$ =100m, enables the grasping of a larger portion of the preferred regions despite the relatively small number of meshes (300 meshes in comparison with 6000, which corresponds to 60 for $\Delta$ of 1 km). In addition, Figure 4 shows the *precision*, which is the percentage of meshes whose centers are within a certain distance from the nearest preferred residential station, among *x* meshes of the highest-MDE for each $\Delta$. Thus, in contrast to the result in Figures 2, the MDE obtained for the smaller $\Delta$ presented a higher performance in detecting preferred residential regions on both criteria of recall and precision, regarded as evidence of well-being of life. This inconsistency will be discussed in the next section.

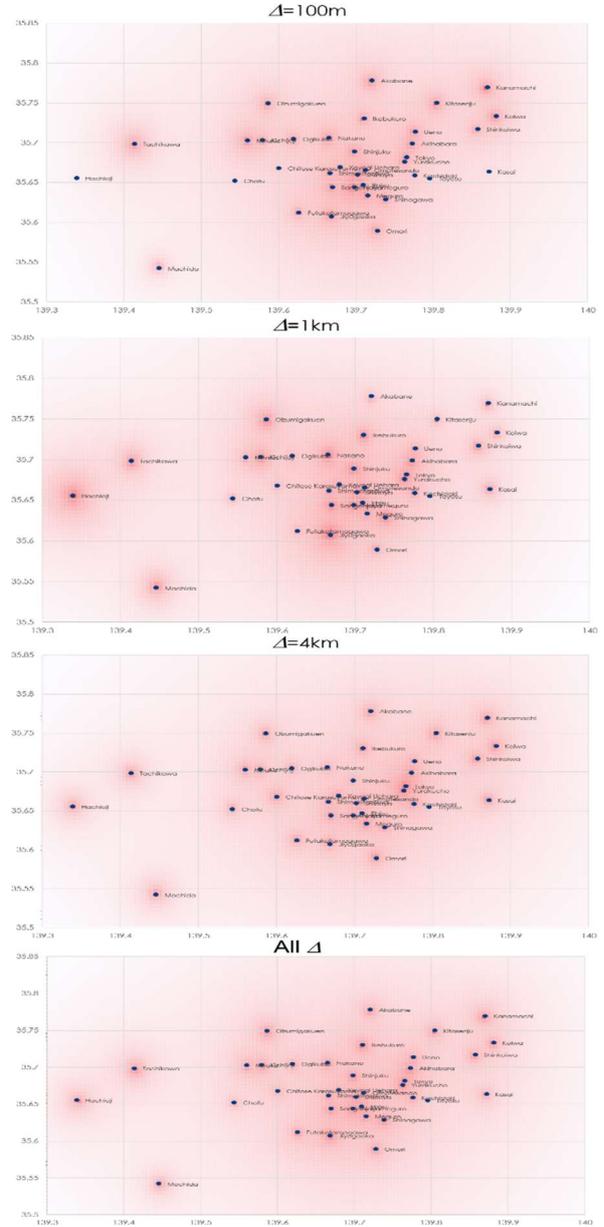

Figure 2 Red-color heatmap for the area (E139.3 – 140.0, N35.5-38.5) showing the high-MDE regions, on which the blue nodes show the 43 most highly preferred stations to live nearby (on three coupled datasets provided by Japanese estate companies [15-17]). The $\Delta$ = all one was the generated by coupling the four results ($\Delta$= 100m, 1km, 2km, and 4km).

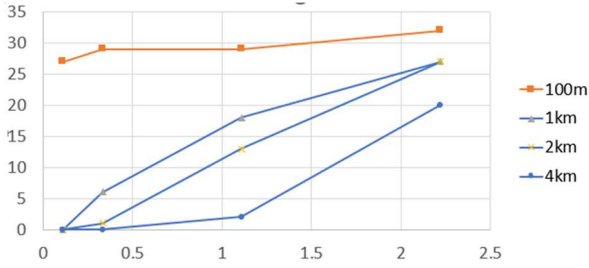

Figure 3 The *recall* performance of MDE. *y*: the number of stations among the 43 in preferred residential regions, within *x* km from the nearest (300 for $\Delta$=100m, 60 for 1km, 60 for 2km, 50 for 4km) grids (centers of meshes) of high MDE.

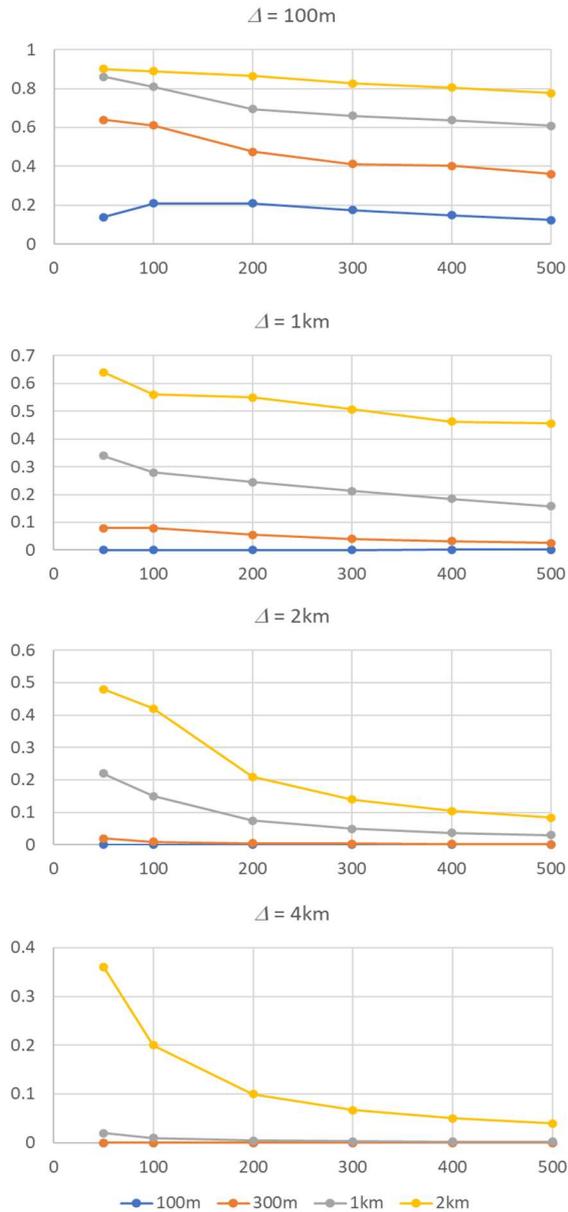

Figure 4 The *precision* of MDE. *y*: the percentage of regions within 100m, 300m, 1km, and 2km from the nearest preferred residential station, among *x* regions of the highest-MDE for each $\Delta$.

## 4. Discussions

The core issue to discuss here is whether we can create a map of regions of well-being using the multi-scale MDE index. In particular, from the results in Section 3, a sub-question is whether a larger $\Delta$ generates a better map, as in Figures 2, or whether the opposite is true, as in Figures 3 and 4.

High-MDE regions at two scaling levels, that is, mesh sizes $\Delta$ of 100m and 4 km, are selectively shown in Figure 5. We found that the high-MDE 100m meshes in the *central* (busiest during daytime, physically in the east) part of Tokyo coincided with regions highly preferred by those exploring places to live in. In contrast to the case $\Delta$ =100m, the meshes of $\Delta$ =4 km, which are in the peripheral area, that is, *out of* the central part, coincide with the preferred regions.

The case of $\Delta$ = 100 m corresponds to the small red mesh in 6 (a) in Figure 6, and the case $\Delta$ =4 km corresponds to the large mesh of the thick red frame surrounding the small blue mesh in (b). In Figure 6, the red color (both frame and shadow) shows the highlights of high-MDE regions, whereas blue shows low-MDE regions. The red frame depicts a large mesh region, in which the low-MDE mesh (b) in the center receives visitors from the surrounding high-MDE regions. That is, the central region imports heterogeneous information from high-activity crowds, which is a crossover of ideas, regardless of the low population density or the low MDE in the central region. The lower part of Figure 6, on the other hand, has a high-MDE mesh in the center (a), where visitors come from various communities, possibly by attending crowds in other regions, to bring heterogeneous ideas. We assume that the central regions in the two types are both crowds of well-being, preferrable to those desiring to live or work in regions of expected future prosperity.

Thus, we found the reason why the precision and recall for $\Delta$ =100m were the largest in Figures 3 and 4, whereas the performance seemed to be the worst, as shown in Figure 2. This is because the central part of Tokyo, where the preferred regions are densely located, was also densely occupied by high-MDE meshes when $\Delta$ =100m as (a) of Figure 6. On the other hand, high-MDE meshes in the case of $\Delta$ = 4 km, as (b) of Figure 6, were concentrated in the peripheral part where the preferred regions were less densely located. However, even in the case of $\Delta$ = 4 km, the local peaks of the MDE in central part of Tokyo were located at the same positions as for $\Delta$ =100m.

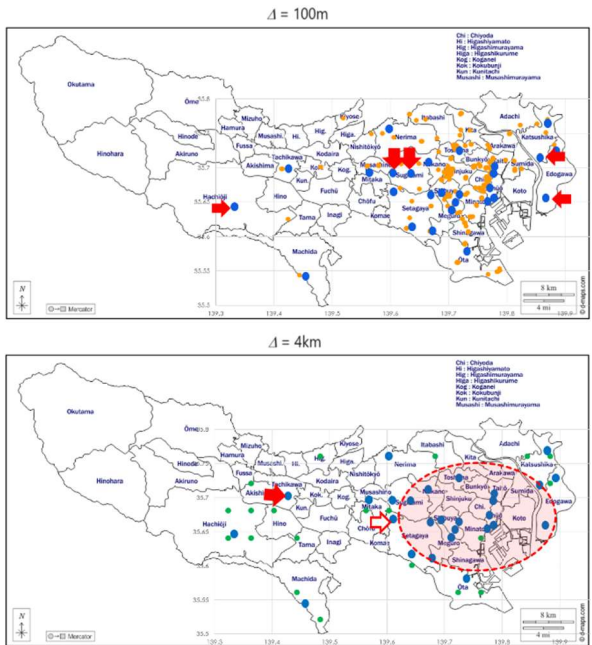

Figure 5 High-MDE meshes by small dots (orange: *Δ*=100m, green: *Δ*=4km). The red arrows and ellipse show regions *not* within distance *Δ* from regions of highly ranked preference to live in according to surveys in [15-17] depicted by the larger dots here. The most active and central metropolitan regions are preferred to live in.

As a result, on the expectation to generate a map of well-being crowds by combining multiscale high-MDE regions from meshes of *Δ* =100m, 1 km, 2 km, and 4 km, we obtained the last one in Figure 2. The correspondence between the peaks of the MDE and the preferred residential regions is visually obvious. Thus, rather than the sheer top-rank MDE regions, which may result only in counting the regions in the highest hills for *Δ*=100m, hill peaks in the heatmap are worth attention in identifying regions preferred to live in.

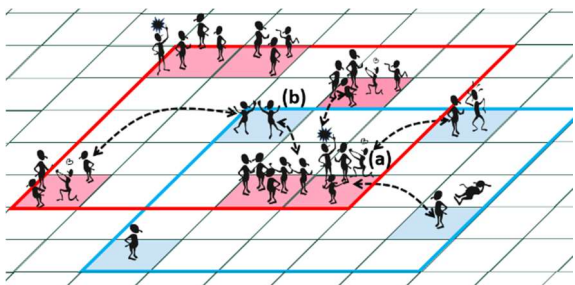

Figure 6 Multiscale high- and low-MDE regions: The red meshes, large or small, show high-MDE regions, whereas the blue show low-MDE ones. The small meshes (a) and (b), red and blue respectively, are the centers of a blue and a red large meshes.

## 5. Conclusions and Future Work

In this study, the well-being of a crowd was interpreted as the crossover of individuals from various communities and quantified using the MDE. The obtained multiscale MDE values reflect a hierarchical social structure involving geographical meshes of various sizes. In this sense, a map of high-MDE regions created by a combination of maps with various granularities can be regarded as visual information about the places where new norms and knowledge are created by the crossover of information from remote regions. Thus, the significant correlation between the maps of preferable residential regions implies a relationship between well-being and creativity.

We will address our future work on the analysis and design of Nigiwai, a Japanese concept often simply translated as a crowd with prosperity. In [18], the measure of the Nigiwai was defined by the physical features of trajectories of human movements and distances within an artificial circumstance, such as a shopping mall. However, the nuance of Nigiwai includes not only the creation of knowledge or norms but also sentimental and emotional enhancement in the place, including interaction with the natural environment. In future work, we aim to extend hierarchical MDE-based generation to consider dimensions other than knowledge.

## Acknowledgment

This study was supported by JST Grant JPMJPF2013 (ClimCore), Q-Leap JPMXS0118067246, JSPS Kakenhi 20K20482, 23H00503, MEXT Initiative for Life Design Innovation, and the Cabinet Secretariat of Japan. The funders played no role in the study design, data collection and analysis, decision to publish, or manuscript preparation. The manuscript has been checked and proofread by Paperpal.

Conflict of Interest declaration: The authors declare that they have NO affiliations with or involvement in any organization or entity with any financial interest in the subject matter or materials discussed in this manuscript.

Author Contributions: YO and MJ contributed to the concepts design of the research, YS and KS to the analysis of the management and analysis of data for the writing of the manuscript.